\documentclass[12pt,preprint]{aastex}
\usepackage{lscape} 

\shorttitle{Eleven new DAVs from the Sloan Survey}
\shortauthors{Mullally et al.}

\begin{document}

\title{Eleven New DAVs from the Sloan Survey}
\author{F. Mullally\altaffilmark{1}, S. E. Thompson\altaffilmark{2}, B. G. Castanheira\altaffilmark{3}, D. E. Winget\altaffilmark{1},  S. O. Kepler\altaffilmark{2}, D. J. Eisenstein\altaffilmark{4}, S. J. Kleinman\altaffilmark{4} and Atsuko Nitta\altaffilmark{4}}
\altaffiltext{1}{Department of Astronomy, 1 University Station, C1400, Austin, TX 78712}
\email{fergal@astro.as.utexas.edu}

\altaffiltext{2}{Department of Physics, Colorado College, 14 E. Cache La Poudre, Colorado Springs, CO 80903}

\altaffiltext{3}{Instituto de F\'{i}sica, Universidade Federal do Rio Grande do Sul, 91501-900 Porto-Alegre, RS, Brazil} 

\altaffiltext{4}{New Mexico State University, Apache Point Observatory, P.O. Box 59, Sunspot, NM 88349}

\begin{abstract}
We report the discovery of eleven new variable DA white dwarf (ZZ Ceti) stars. Candidates were selected by deriving temperatures from model fits to spectra obtained from the Sloan Digital Sky Survey. We also find objects whose temperatures and gravities indicate they lie within the instability strip for
pulsation, but which were not observed to vary. Although the temperatures are based on relatively low S/N spectra, an impure strip is unexpected, and if confirmed suggests that our knowledge of the pulsation mechanism is incomplete. This work brings the total number of published variable DA white dwarf stars to 82.

\end{abstract}


\section{Introduction}
The relatively simple structure and behavior of white dwarf stars (WDs) make them ideal objects for astrophysical study. 
For the variable WDs, asteroseismology allows us a rare glimpse into the interior of a stellar object. WDs pulsate in three distinct instability strips along the HR diagram. The extremely high gravity of these objects makes  non-radial gravity-modes energetically favorable \citep[and references therein]{Winget98}. Of interest in this paper are the hydrogen atmosphere WDs (known as the DAs) which pulsate at temperatures between approximately $11,000$~K and $12,000$~K \citep{Mukadam04}. We previously believed that variability was a normal part of the evolution of a cooling white dwarf \citep{Fontaine82, Bergeron04}, so these pulsating WDs (or DAVs) are otherwise normal stars caught during the brief period of evolution where their temperatures allow pulsation.  
 However recent analysis by \citet{Mukadam04b} has shown the presence of non-variable stars within the strip, indicating either that the models used for fitting temperatures need refinement, or the presence of an additional third parameter determining the pulsation properties of these objects. This is an important concern in the application of the conclusions of DA asteroseismology to other DAs.

A hot subset of the variable DAs (known as hDAVs) were discovered to exhibit extreme stability in the period and phase of their pulsations \citep{Stover80, Kepler82}. 
Kepler \citetext{in press} showed that one such star, G117-B15A has a period stability of $\dot{P}=(4.12\pm0.83)\times10^{-15}$, a stability that rivals that of atomic clocks. \citet{Mukadam03a} constrained the stability of ZZ Ceti to better than $(5.5\pm1.9)\times10^{-15}$. 

With such a stable signal the presence of an orbiting planet can be inferred from variations in the observed arrival time of pulsations due to the reflex orbital motion of the star. The first limits on planetary mass companions to white dwarf stars were placed by \citet{Kepler88}. For this paper, our search for new variables was biased toward the hot edge of the strip where these stable pulsators, suitable for searching for planets, are to be found.

A key constraint on both the prior progress of asteroseismology and the search for planetary companions was the limited number of suitable stars available for study. For that reason, \citet{Mukadam04} performed a photometric search and discovered 35 new DAVs. This search is on-going, and in this paper we report 11 new stars to make them available to the wider community.  We refer the reader to \citet{Mukadam04} for a full description of this program.

\section{Object Selection and Observation}
The Sloan Digitial Sky Survey \citep{Fukugita96, Gunn98, York00, Hogg01, Smith02, Stoughton02, Pier03} is proving to be an impressive source of new white dwarf stars \citep{Kleinman04}. We obtained candidate DAVs from both the DR1 \citep{Abazajian03} and DR2 \citep{Abazajian04} samples. Objects from DR1 were selected from the catalogue of \citet{Kleinman04} using temperature fits based on models published in \citet{Finley97}. 

Objects from DR2 \citep[which do not appear in][]{Kleinman04} with spectra were also selected. DA stars near the DA instability strip are easily 
identifable due to their very broad Balmer lines caused by their very high surface gravity and the fact that the Balmer lines are maximally broad near the temperature range of the instability strip \citep{Fontaine03}. For each spectrum in the database we measured the equivalent widths of the $H_{\beta}$ and $H_{\gamma}$ lines over the wavelength region given in Table \ref{ew}. Objects in the range 40\AA~$ < H_{\beta} < $65\AA~and 20\AA~$ < H_{\gamma} <$ 45\AA~were selected and a colour cut of $0.2 \leq (u-g) \leq 0.7$, $-0.4 \leq (g-r) \leq 0.05$ and $9.5(u-g) - (g-r) > 4.14$ was used to further trim the sample. The third cut removes DAs with Balmer lines of appropriate equivalent width but on the hotter side of the curve of growth ($\approx$ 15,000~K). The temperatures and gravities of the selected DAVs were found by fitting to a grid of temperature models as described in \citet{Kleinman04}

Objects were observed and reduced as described in \citet{Mukadam04}. Each 
object was observed for two hours on the 2.1m Otto Struve telescope at McDonald Observatory using the Argos prime focus CCD camera \citep{Nather04}. Individual exposure times were between 5 and 15 seconds depending on the brightness of the target and readout times were negligible due to the use of a frame transfer buffer. If an object showed signs of variability it was reobserved on a later night for confirmation. Faint objects or those observed under poor conditions may appear to show variability, so a second run is required to confirm variability. If an object did not appear to pulsate it was not reobserved. Many DAVs present closely spaced modes which can destructively interfere effectively hiding a mode for periods longer than two hours. However the aim of this survey is to find as many pulsators as possible with the telescope time available not to conduct a complete search of the sample and so stars that did not appear to vary were not re-observed.

The CCD images were flat fielded and lightcurves extracted using IRAF's weighted aperture appphot package. We subtract the contribution from sky photons and divide by a combination of reference stars to remove small cloud variations.

\begin{figure}[tbh]
    \begin{center}
        \scalebox{.5}{\rotatebox{270}{\includegraphics{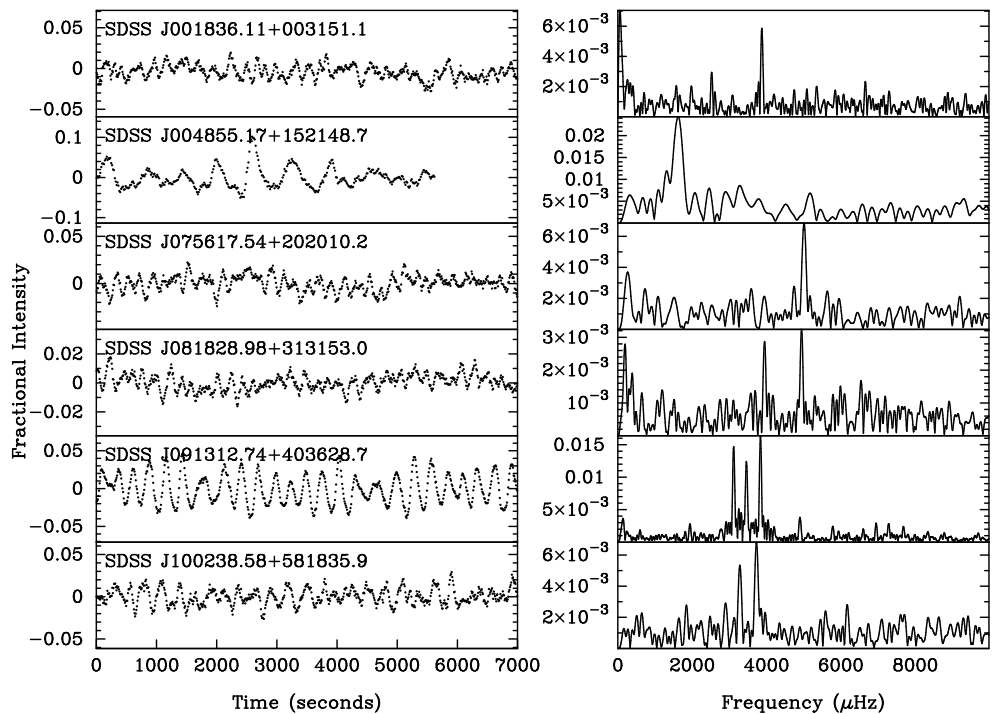}}}
        \caption[mull04a]{ Two hour portions of lightcurves for the new pulsators. The lightcurves have been boxcar smoothed by seven points to emphasize the pulse shapes. The Fourier transforms in the right column are of the unsmoothed data and may be taken from longer datasets. \label{mull04a}}
    \end{center}
\end{figure}

\begin{figure}[bth]
    \begin{center}
        \scalebox{.5}{\rotatebox{270}{\includegraphics{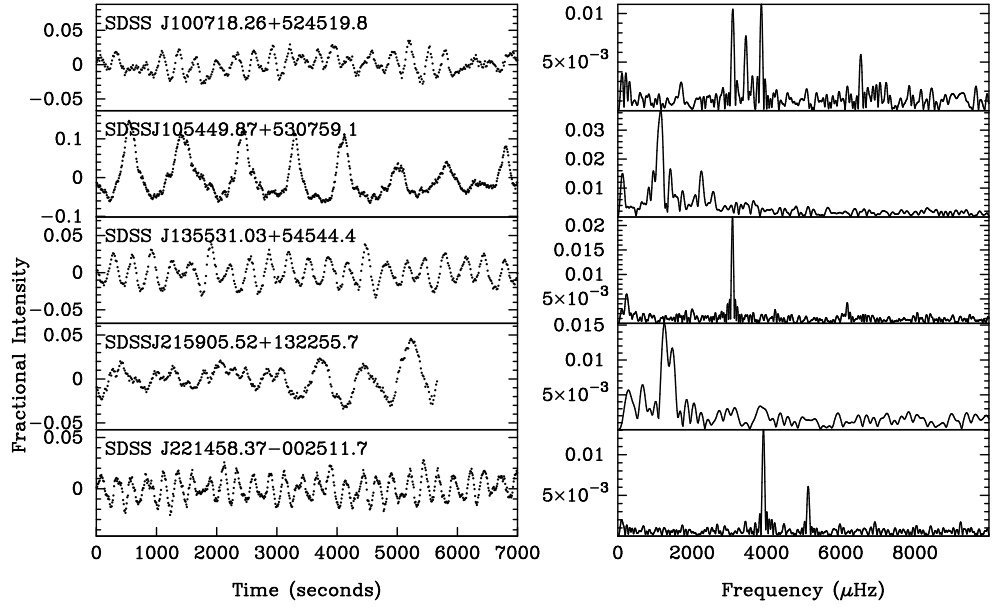}}}
        \caption[mull04b]{Same as Figure~\ref{mull04a} for five additional pulsators. \label{mull04b}}
    \end{center}
\end{figure}

We discovered 11 new DAVs and 26 stars that were not observed to vary. A journal of observations appears in Table~\ref{journaltab}. Lightcurves and Fourier transforms of the new pulsators are shown in Figures~\ref{mull04a} and \ref{mull04b}. Table~\ref{periodstab} lists the observed periods and amplitudes of pulsation. The high number of non-variables is due to an unsuccessful attempt to use a different method to measure the stellar temperatures. Our instability strip is similar to that discovered in \citet{Mukadam04}, which is to be expected as we are using the same temperature fitting technique.

\section{Characteristics of the Instability Strip}

\begin{figure}[p]
    \begin{center}
        \scalebox{.5}{\rotatebox{270}{\includegraphics{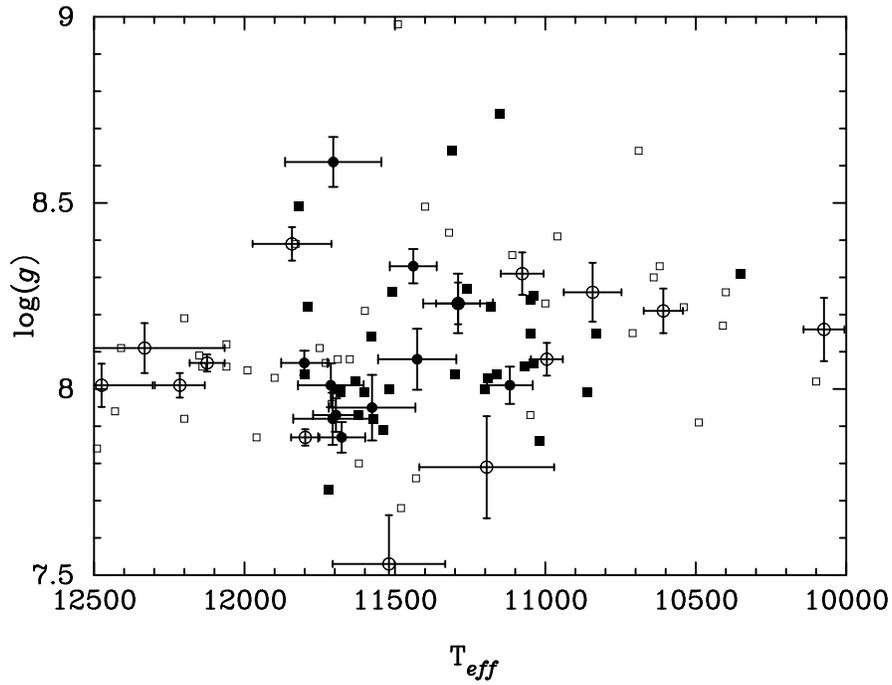}}}
        \caption[]{Distribution of effective temperatures and gravities of DAVs discovered in the Sloan survey. The filled shapes are pulsators, hollow shapes are NOVs. Circles are stars reported in this paper while squares are from \citet{Mukadam04}. For clarity errorbars are only shown for objects reported in this paper, those in \citet{Mukadam04} are similar in size.  \label{tvg}}
    \end{center}
\end{figure}

A plot of the location of the new variables within the instability strip is shown in Figure~\ref{tvg}. A table of the properties of the variables is presented in Table~\ref{davtab} and those stars not observed to vary in Table~\ref{novtab}. The phrase non-variable is fraught with danger, as a star may be exhibiting destructive interference between two closely spaced modes while being observed, or merely be pulsating with too low an amplitude to be detected. For this reason, we prefer to use the term Not Observed to Vary (NOV).

The uncertainties in Figure~\ref{tvg} and Tables~\ref{davtab} and \ref{novtab} are the formal least squares fit errors. In an effort to determine the extent of external errors in DAV temperature fits, \citet{Fontaine03} compared the measured effective temperature of a number of DAs in the region of the instability strip as measured from two independently observed and reduced spectra of each object. They conclude that the external errors, due primarily to different flux calibrations, was $\sim$200~K. It should be noted that paper uses spectra with signal to noise ratios of greater than $80$ per pixel, while our faintest star, SDSS J173712 ($g=19.2$) has a S/N ratio of less than $8$. \citet{Mukadam04b}, using similar spectra from the Sloan survey which are observed and reduced in a consistent manner, estimate an uncertainty in $T_{eff}$ of less than $300$~K for the fainter stars, and $200$~K for the brighter stars.

Our enlarged sample of DAVs has the same characteristics as the sample published in \citet{Mukadam04}. Our survey emphasized the blue edge of the instability strip which is why we found more pulsators hotter than $11,500$~K than cooler. With this bias in mind, our new sample still supports the narrower strip found in \citet{Mukadam04b}. We note that two stars not observed to vary, SDSS J143249 and SDSS J012234 lie within the strip. It is possible that these objects are complex pulsators whose modes were destructively interfering for the time they were observed, or simply that their amplitude was too low to be observed. If further observations address those concerns, these objects lend support to the arguments in \citet{Mukadam04b} that the DA instability strip is impure.

\acknowledgments
This work is supported by a grant from the NASA Origins program, NAG5-13094 and performed in part under contract with the Jet Propulsion Laboratory (JPL) funded by NASA through the Michelson Fellowship Program. JPL is managed for NASA by the California Instituute of Technology. We also acknowledge the support of the Texas Advanced Research Program under grant ARP-0543.

    Funding for the creation and distribution of the SDSS Archive has been provided by the Alfred P. Sloan Foundation, the Participating Institutions, the National Aeronautics and Space Administration, the National Science Foundation, the U.S. Department of Energy, the Japanese Monbukagakusho, and the Max Planck Society. The SDSS Web site is \url{http://www.sdss.org/}.

The SDSS is managed by the Astrophysical Research Consortium (ARC) for the Participating Institutions. The Participating Institutions are The University of Chicago, Fermilab, the Institute for Advanced Study, the Japan Participation Group, The Johns Hopkins University, the Korean Scientist Group, Los Alamos National Laboratory, the Max-Planck-Institute for Astronomy (MPIA), the Max-Planck-Institute for Astrophysics (MPA), New Mexico State University, University of Pittsburgh, Princeton University, the United States Naval Observatory, and the University of Washington.

We thank D. Schneider for his criticisms of an early draft of this manuscript.


\begin{table}[htb]
\begin{center}
\caption{Wavelengths used to calculate equivalent width of Balmer lines \label{ew}}
\begin{tabular}{rrr}
\tableline
\tableline
Line& Center (\AA)& Width (\AA)\\
\tableline
$H_{\beta}$ & $ 4861.3$ & $ 324$\\
$H_{\gamma}$ & $ 4340.5$ & $ 214$\\
\tableline
\end{tabular}
\end{center}
\end{table}

\begin{deluxetable}{cccccc}
\tabletypesize{\footnotesize}
\tablewidth{0pc}
\tablecaption{Journal of Observations \label{journaltab}}
\tablehead{
	\colhead{Run}& \colhead{ObjectName}& \colhead{UTC Date}&
	\colhead{Start time}& \colhead{Exp}& \colhead{Length}\\
}

\startdata
A0752& SDSS J001836.11$+$003151.1& 2003-11-19& 03:05:56& 15& 02:05:45 \\
A0762& SDSS J001836.11$+$003151.1& 2003-11-21& 04:18:46& 15& 02:05:00 \\
A0794& SDSS J001836.11$+$003151.1& 2003-12-01& 00:55:45& 10& 04:42:00 \\
A0701& SDSS J004855.17$+$152148.7& 2003-09-04& 08:10:40& 15& 01:40:00 \\
A0706& SDSS J004855.17$+$152148.7& 2003-09-05& 09:09:51& 15& 01:52:15 \\
A0860& SDSS J075617.54$+$202010.2& 2004-03-18& 01:54:49& 10& 03:08:20 \\
A0864& SDSS J075617.54$+$202010.2& 2004-03-19& 02:03:04& 10& 04:55:00 \\
A0831& SDSS J081828.98$+$313153.0& 2004-01-19& 03:57:18& 10& 03:26:40 \\
A0849& SDSS J081828.98$+$313153.0& 2004-03-01& 01:58:28& 10& 02:51:20 \\
A0836& SDSS J091312.74$+$403628.7& 2004-01-20& 07:43:18& 10& 01:44:20 \\
A0866& SDSS J091312.74$+$403628.7& 2004-03-20& 02:03:26& 10& 05:00:10 \\
A0861& SDSS J100238.58$+$581835.9& 2004-03-18& 05:11:27& 10& 03:06:40 \\
A0870& SDSS J100238.58$+$581835.9& 2004-03-24& 05:05:08& 10& 01:23:00 \\
A0635& SDSS J100718.26$+$524519.8& 2003-05-06& 02:42:38& 15& 03:51:15 \\
A0869& SDSS J100718.26$+$524519.8& 2004-03-24& 02:01:27& 15& 01:45:45 \\
A0833& SDSS J105449.87$+$530759.1& 2004-01-19& 09:45:01& 10& 03:19:30 \\
A0867& SDSS J105449.87$+$530759.1& 2004-03-20& 07:14:37& 10& 02:45:50 \\
A0862& SDSS J135531.03$+$545404.5& 2004-03-18& 08:25:51& 15& 02:03:00 \\
A0873& SDSS J135531.03$+$545404.5& 2004-03-25& 09:25:52& 15& 02:43:00 \\
A0880& SDSS J135531.03$+$545404.5& 2004-05-14& 02:58:42& 15& 05:09:00 \\
A0430& SDSS J215905.52$+$132255.7& 2002-12-08& 00:53:08& 15& 01:53:00 \\
A0673& SDSS J215905.52$+$132255.7& 2003-07-02& 09:22:30& 15& 01:43:45 \\
A0670& SDSS J221458.37$-$002511.7& 2003-07-01& 08:37:02& 10& 02:36:20 \\
A0692& SDSS J221458.37$-$002511.7& 2003-09-02& 04:35:53& 10& 03:59:50 \\
A0723& SDSS J221458.37$-$002511.7& 2003-10-25& 01:15:40& 10& 04:18:30 \\
A0761& SDSS J221458.37$-$002511.7& 2003-11-21& 00:54:57& 10& 03:15:30 \\
A0783& SDSS J221458.37$-$002511.7& 2003-11-28& 01:01:22& 10& 02:55:00 \\
\enddata
\end{deluxetable}

\begin{deluxetable}{lcccc}
\tabletypesize{\footnotesize}
\tablewidth{0pc}
\tablecaption{Observed Periods and Amplitudes \label{periodstab}}
\tablehead{
    \colhead{Object} &\colhead{Resolution}& \colhead{Frequency }& 
	\colhead{Period }& \colhead{Amplitude }\\

	\colhead{~ } & \colhead{($\mu$Hz)} & \colhead{($\mu$Hz)}& 
	\colhead{(sec)}& \colhead{(\%)}}

\startdata
SDSS J001836.11$+$003151.1& ~59 &3876~  & 257.9 & 0.58\\[1ex]
SDSS J004855.17$+$152148.7& 145 &1625$^*$& 615.3 & 2.48\\[1ex]
SDSS J075617.54$+$202010.2& ~56 &5011~ & 199.5 & 0.68\\[1ex]
SDSS J081828.98$+$313153.0& ~81 &3947$^*$& 253.3 & 0.29\\
 &  & 4942~ & 202.3 & 0.33\\[1ex]
SDSS J091312.74$+$403628.7& ~56 &3119$^*$& 320.5 & 1.47\\
 &  & 3462~ & 288.7 & 1.24\\
 &  & 3841$^*$& 260.3 & 1.65\\
 &  & 4903~ & 203.9 & 0.38\\[1ex]
SDSS J100238.58$+$581835.9& ~89 &3282~ & 304.6 & 0.53\\
 &  & 3728~ & 268.2 & 0.68\\[1ex]
SDSS J100718.26$+$524519.8& ~72 &3094$^*$& 323.1 & 1.04\\
 &  & 3446~ & 290.1 & 0.77\\
 &  & 3863$^*$& 258.8 & 1.10\\
 &  & 6540~ & 152.8 & 0.58\\[1ex]
SDSS J105449.87$+$530759.1& 101 &1150$^*$& 869.1 & 3.74\\
 &  & 2248~ & 444.6 & 1.60\\[1ex]
SDSS J135531.03$+$545404.4& ~54&3086~ & 324.0 & 2.18\\[1ex]
SDSS J215905.52$+$132255.7& 147&1248~ &801.0 &1.51\\
 &  & 1462$^*$& 683.7 & 1.17\\[1ex]
SDSS J221458.37$-$002511.7& ~65&3917~ & 255.2 & 1.31\\
 &  & 5122~ & 195.2 & 0.61\\
\enddata

\tablecomments{We do not have the resolution in our datasets to resolved multiplets or closely spaced modes for most of these stars.  Objects marked $^*$ show evidence of amplitude variablity between runs. The resolution quoted is reciprocal of the length of the run. 
}
\end{deluxetable}

\begin{landscape}
\begin{table}[htb]
\begin{footnotesize}
\caption{Properties of new DAVs \label{davtab}}
\begin{tabular}{rrrlr@{$\pm$}lr@{$\pm$}lr@{$\pm$}lr@{$\pm$}lrrr}
\tableline
\tableline

Mjd&Plate&Fiber&Designation& \multicolumn{2}{c}{$T_{eff}$} &\multicolumn{2}{c}{log($g$)} &\multicolumn{2}{c}{$H_{\beta}$}&\multicolumn{2}{c}{$H_{\gamma}$}&$u - g$&$g - r$&$g$\\

\tableline

52203& 0688& 348& SDSS J001836.11$+$003151.1& 11696& 076& 7.93& 0.045& 52.66& 0.81& 32.70& 0.59& 0.452& -0.160& 17.360\\
51871& 0420& 388& SDSS J004855.17$+$152148.7& 11290& 116& 8.23& 0.080& 52.51& 1.58& 37.41& 1.13& 0.401& -0.109& 18.676\\
52941& 1583& 167& SDSS J075617.54$+$202010.2& 11713& 109& 8.01& 0.059& 54.92& 1.33& 36.07& 0.98& 0.465& -0.150& 18.240\\
52619& 0931& 321& SDSS J081828.98$+$313153.0& 11801& 077& 8.07& 0.033& 55.25& 0.80& 38.65& 0.57& 0.382& -0.185& 17.381\\
52668& 1200& 017& SDSS J091312.74$+$403628.8& 11677& 078& 7.87& 0.041& 53.90& 1.02& 37.78& 0.71& 0.495& -0.224& 17.635\\
52317& 0558& 573& SDSS J100238.58$+$581835.9& 11707& 131& 7.92& 0.070& 54.82& 1.32& 36.83& 0.95& 0.480& -0.201& 18.264\\
52400& 0903& 557& SDSS J100718.26$+$524519.8& 11426& 130& 8.08& 0.082& 56.47& 1.68& 37.61& 1.19& 0.414& -0.162& 18.872\\
52649& 1010& 629& SDSS J105449.87$+$530759.1& 11118& 076& 8.01& 0.050& 50.81& 1.02& 37.85& 0.71& 0.451& -0.156& 17.922\\
52797& 1323& 161& SDSS J135531.03$+$545404.5& 11576& 144& 7.95& 0.088& 52.32& 1.48& 36.70& 1.05& 0.398& -0.146& 18.583\\
52224& 0734& 419& SDSS J215905.52$+$132255.7& 11705& 160& 8.61& 0.067& 54.86& 1.75& 37.52& 1.28& 0.381& -0.193& 18.873\\
51791& 0374& 180& SDSS J221458.37$-$002511.7& 11439& 078& 8.33& 0.046& 52.41& 1.06& 36.63& 0.77& 0.334& -0.099& 17.909\\
\tableline
\end{tabular}
\end{footnotesize}
\end{table}
\end{landscape}

\begin{landscape}
\begin{table}
\begin{footnotesize}
\caption{Table of objects not observed to vary \label{novtab}}
\begin{tabular}{rrrlc@{$\pm$}cr@{$\pm$}lr@{$\pm$}lr@{$\pm$}llllc}
\tableline
\tableline
\renewcommand{\arraystretch}{.75}
Mjd&Plate&Fiber&Designation& \multicolumn{2}{c}{$T_{eff}$} &\multicolumn{2}{c}{log($g$)} &\multicolumn{2}{c}{$H_{\beta}$}&\multicolumn{2}{c}{$H_{\gamma}$}&$u - g$&$g - r$&\multicolumn{1}{c}{$g$}&Limit (mma)\\
\tableline

51900 & 0390 & 455 & SDSS J002049.39$+$004435.0 & 9160 & 10 & 9.00 & 0.003 & 25.60 & 0.74 & 15.47 & 0.55 & 0.238 & 0.058 & 16.797&1\\
52203 & 0688 & 164 & SDSS J002309.03$-$003342.0 & 15522 & 81 & 8.01 & 0.016 & 55.56 & 0.49 & 38.94 & 0.34 & 0.242 & -0.300 & 16.280&1\\
51879 & 0419 & 098 & SDSS J004610.37$+$133910.2 & 11077 & 71 & 8.31 & 0.057 & 52.31 & 1.24 & 33.07 & 0.90 & 0.404 & -0.132 & 18.040&2\\
51871 & 0420 & 591 & SDSS J005703.73$+$151014.6 & 10074 & 68 & 8.16 & 0.085 & 43.11 & 1.76 & 23.76 & 1.32 & 0.546 & -0.063 & 18.850&4\\
52209 & 0696 & 476 & SDSS J012234.67$+$003026.3 & 11798 & 47 & 7.87 & 0.022 & 55.66 & 0.61 & 34.94 & 0.42 & 0.355 & -0.073 & 17.286&2\\
52178 & 0702 & 448 & SDSS J020851.65$+$005332.4 & 13401 & 150 & 7.77 & 0.024 & 58.75 & 0.64 & 38.40 & 0.44 & 0.386 & -0.270 & 16.960&2\\
51869 & 0406 & 385 & SDSS J022108.67$+$004924.7 & 10608 & 65 & 8.21 & 0.060 & 46.24 & 1.57 & 35.73 & 1.11 & 0.457 & -0.124 & 18.632&2\\
51816 & 0410 & 501 & SDSS J025709.00$+$004628.0 & 12215 & 83 & 8.01 & 0.033 & 57.51 & 0.84 & 38.97 & 0.60 & 0.418 & -0.206 & 17.387&2\\
52203 & 0710 & 548 & SDSS J031111.38$-$000344.4 & 14537 & 197 & 8.32 & 0.040 & 63.66 & 0.93 & 37.63 & 0.66 & 0.295 & -0.200 & 17.870&3\\
51929 & 0413 & 074 & SDSS J032302.85$+$000559.6 & 13030 & 158 & 7.98 & 0.041 & 59.63 & 0.89 & 42.24 & 0.62 & 0.785 & -0.220 & 17.436&4\\
51901 & 0414 & 273 & SDSS J032510.84$-$011114.1 & 18267 & 86 & 7.59 & 0.017 & 46.27 & 0.74 & 30.97 & 0.55 & 0.446 & -0.191 & 17.073&3\\
51901 & 0414 & 454 & SDSS J032619.44$+$001817.5 & 12124 & 58 & 8.07 & 0.023 & 58.08 & 0.87 & 38.25 & 0.61 & 0.387 & -0.207 & 17.420&6\\
51810 & 0415 & 206 & SDSS J033200.49$-$005752.5 & 17476 & 109 & 7.77 & 0.023 & 45.24 & 0.78 & 32.68 & 0.54 & 0.227 & -0.299 & 17.062&2\\
52370 & 0771 & 225 & SDSS J101218.09$+$610818.9 & 11842 & 131 & 8.39 & 0.045 & 55.39 & 1.01 & 36.78 & 0.74 & 0.399 & -0.188 & 17.733&2\\
52378 & 0838 & 144 & SDSS J114132.99$+$042028.8 & 11520 & 187 & 7.53 & 0.131 & 49.85 & 1.31 & 36.37 & 0.94 & 0.524 & -0.222 & 18.186&5\\
51984 & 0498 & 234 & SDSS J140004.68$+$643128.3 & 10995 & 53 & 8.08 & 0.044 & 48.85 & 1.05 & 36.05 & 0.74 & 0.467 & -0.206 & 17.671&3\\
52024 & 0536 & 318 & SDSS J143249.11$+$014615.5 & 11290 & 73 & 8.23 & 0.056 & 52.60 & 0.96 & 38.12 & 0.70 & 0.540 & -0.170 & 17.484&2\\
52045 & 0594 & 478 & SDSS J154545.35$+$032150.0 & 15652 & 272 & 7.97 & 0.054 & 54.98 & 1.67 & 36.45 & 1.19 & 0.329 & -0.242 & 18.753&3\\
52395 & 0818 & 476 & SDSS J164248.61$+$382411.1 & 18813 & 204 & 8.40 & 0.035 & 51.50 & 1.20 & 34.64 & 0.83 & 0.076 & -0.322 & 17.952&3\\
52438 & 0820 & 516 & SDSS J165815.53$+$363816.0 & 10843 & 96 & 8.26 & 0.079 & 50.22 & 1.98 & 30.05 & 1.46 & 0.455 & -0.119 & 19.162&5\\
52017 & 0366 & 629 & SDSS J173712.95$+$584428.7 & 11195 & 224 & 7.79 & 0.137 & 49.37 & 2.14 & 37.36 & 1.50 & 0.517 & -0.182 & 19.266&3\\
52224 & 0734 & 348 & SDSS J215532.95$+$123801.5 & 12332 & 266 & 8.11 & 0.067 & 55.03 & 1.34 & 38.07 & 0.97 & 0.376 & -0.223 & 18.338&2\\
52518 & 0737 & 226 & SDSS J222223.04$+$123824.7 & 13888 & 142 & 7.58 & 0.030 & 53.84 & 0.87 & 40.60 & 0.59 & 0.388 & -0.235 & 17.63&1\\
52263 & 0740 & 601 & SDSS J225211.51$+$143610.5 & 17027 & 219 & 7.86 & 0.044 & 49.09 & 1.25 & 32.85 & 0.87 & 0.213 & -0.286 & 18.148&2\\
52251 & 0744 & 273 & SDSS J231152.20$+$142417.2 & 12475 & 170 & 8.01 & 0.058 & 57.03 & 1.30 & 39.03 & 0.95 & 0.448 & -0.227 & 18.251&3\\
52553 & 0647 & 126 & SDSS J233742.20$-$104144.0 & 16164 & 109 & 7.89 & 0.024 & 52.29 & 0.79 & 36.73 & 0.54 & 0.168 & -0.293 & 17.188&3\\

\tableline
\end{tabular}
\end{footnotesize}
\end{table}
\end{landscape}


\end{document}